# Dual-polarization multiplexing amorphous Si:H grating couplers for silicon photonic transmitters in the photonic BiCMOS backend of line

Galina GEORGIEVA[1,*], Christian MAI[2], Pascal M. SEILER[1,2], Anna PECZEK[3] and Lars ZIMMERMANN[1,2]

1 Technische Universität Berlin, Hochfrequenztechnik-Photonik/Siliziumphotonik, Straße des 17. Juni 135, 10623 Berlin, Germany

2 IHP – Leibniz Institut für innovative Mikroelektronik, Im Technologiepark 25, 15236 Frankfurt (Oder), Germany

3 IHP Solutions GmbH, Im Technologiepark 7, 15236 Frankfurt (Oder), Germany

* Corresponding author: galina.georgieva@tu-berlin.de

**Abstract** We report on polarization combining 2D grating couplers (2D GCs) on amorphous Si:H, fabricated in the backend of line of a photonic BiCMOS platform. The 2D GCs can be used as an interface of a hybrid silicon photonic coherent transmitter, which can be implemented on bulk Si wafers. The fabricated 2D GCs operate in the telecom C-band and show an experimental coupling efficiency of -5 dB with a wafer variation of ±1.2 dB. Possibilities for efficiency enhancement and improved performance stability in future design generations are outlined and extension towards O-band devices is investigated as well.

## 1 Introduction

In the optical communications community, there is presently an extensive discussion on which technology will enable high bit rates for data center interconnects in the near future [1]. Coherent transceivers already started entering commercially the data center domain, e.g. [2], [3]. Considering even shorter distances, there are two crucial factors for the ultimate establishment of coherent formats: cost and power dissipation. With respect to cost, silicon (Si) photonics is an attractive platform, which offers a highly scalable technology and a monolithic co-integration of photonics and electronics, by using mature complementary metal-oxide-semiconductor (CMOS) foundry processes [4], [5], [6]. There is one significant obstacle on the way of coherent silicon photonic solutions for data centers, which is the silicon modulator and the modulator drivers [7]. Due to the fundamental limitations of all-silicon phase shifters, silicon Mach-Zehnder modulators frequently show opto-electrical bandwidth < 50 GHz, large modulation loss and $V_\pi$ and considerable power consumption [3], [8], [9]. For these reasons, many groups started considering the accomplishment of silicon hybrid transmitters, in which the modulator is made of material with distinct second-order nonlinearities (Pockels effect). Modulators based on various materials have been reported, such as lead zirconate titanate [10], barium titanate [11] and lithium niobate [12], [13], [14]. Moreover, attempts towards the co-integration on a full-flow silicon platform have been undertaken [11], [14]. All Pockels materials listed here are not CMOS compatible and require a post fabrication bonding step. From the perspective of a foundry process flow, bonding of a Pockels material in the frontend of line (FEOL) would require significant modifications of the FEOL fabrication routine, adding processing time and complexity. In comparison, bonding after the backend of line (BEOL) has been processed is less challenging and both IBM [11] and Sandia [14] demonstrated their modulators in the BEOL. If we consider keeping the modulator there, an extension with another photonic layer at that level is necessary. This leads us to the concept of the three-dimensional (3D) photonic integration, which attracted attention in previous years due to its improved integration density and higher components design flexibility [15]. The extension to more than one photonic layers can be achieved by using CMOS process compatible materials such as hydrogenated amorphous silicon (a-Si:H) [16], silicon nitride ($Si_3N_4$) [17], [18] or aluminum nitride (AlN) [19]. Among them, a-Si:H offers the highest refractive index contrast, which makes it best suitable for inter- or intra-chip vertical grating coupling [20]. The initial disadvantage of the high absorption loss of a-Si at telecom wavelengths could be overcome by hydrogen passivation [21]. For that reason, a-Si:H is a promising material for 3D photonic transmitters or transceivers.

Being able to define a photonic layer in the BEOL and to assemble a Pockels modulator at the same level, as already demonstrated in references [11], [14], we would need only an in- and out-coupling interface for the accomplishment of the transmitter's photonic part. Here, we focus on the realization of dual-polarization coupling structures, which are intended as a future part of hybrid silicon photonic coherent transmitters in a photonic BiCMOS BEOL. For that purpose, we consider two-dimensional grating couplers (2D GCs) best suitable, because they allow for a polarization-multiplexed transmission without the necessity of an additional polarization rotator. Furthermore, their large mode spot size is an important advantage with respect to packaging tolerances and cost, which are of great importance in the context of coherent formats for data center interconnects.

In this work, a-Si:H waveguides and grating couplers are realized at the Metal2 level of a photonic BiCMOS platform [22], [23]. The platform combines bulk Si electronics and silicon-on-insulator (SOI) photonics in the FEOL that share a common BEOL. The a-Si:H structures are intended as a proof-of-concept and are used for the investigation of material and process parameters as well. In a future generation, an integration at the TopMetal1 level will be pursued. This is motivated by the successful demonstration of a Pockels modulator at the same level on this platform [11]. Another benefit is the possibility to improve the efficiency and the performance stability of the a-Si:H 2D GCs by incorporating the BEOL Metal3 as a reflector. For comparison, a comprehensive review on the coupling to crystalline Si photonics may be found in [24].

The paper is structured as follows: section 2 outlines the platform concept, in which 2D GCs will be necessary, section 3 gives details on the numerical investigation of 2D GCs with two considered scenarios – structures at Metal2 for experimental comparison and structures at TopMetal1 for future optimization and co-integration with other devices. Section 4 describes the fabrication process and its evaluation in terms of material and geometric variations. Section 5 shows experimental results on the first generation of fabricated a-Si:H waveguides and grating couplers at the Metal2 level, including waveguide loss and grating coupler efficiency. The grating couplers are compared to the same designs, realized on SOI. The final section 6 summarizes the results and gives an outlook for our future work.

## 2 Concept

In this section, we outline our general concept for a dual-polarization coherent transmitter in a photonic BiCMOS BEOL. Fig. 1 illustrates our photonic BiCMOS platform [22], [23], extended by a second photonic a-Si:H layer near the TopMetal1 level. The combination of the a-Si:H layer with a Pockels material of choice can enable the realization of high-speed transmitters, which meet the requirements for future generation data center interconnects. The separation of the metal levels, where Pockels materials and a-Si:H are integrated, requires a thick enough interlayer stack. The BEOL of our photonic BiCMOS technology consists of three thin (Metal1, Metal2 and Metal3) and two thick (TopMetal1 and TopMetal2) metal layers. The interlayer dielectric stacks above the thin metal layers have a thickness of around 900 nm, which is not sufficient. For that reason, the integration of a-Si:H after processing TopMetal1 is favorable, since the interlayer dielectric stack between TopMetal1 and TopMetal2 with a thickness of around 3 µm fulfills best the requirements.

In the photonic a-Si:H layer, typical components such as waveguides, directional couplers, multi-mode interferometers (MMI) etc. can be realized. Grating couplers can be used as in- and out-coupling interfaces, whereat the input interface can be substituted in future by a hybrid integrated laser source, following the principles outlined e.g. in references [25], [26], [27]. For polarization-multiplexed systems, 2D GCs are necessary as an output. Regarding the efficiency of grating couplers, a low thickness variation of silicon dioxide ($SiO_2$) is very decisive. There are several process points in the BEOL, which cause variation in the $SiO_2$ cladding. Before the fabrication of each metal layer, a chemo-mechanical polishing (CMP) is done to create suitable process conditions. This means that we have four CMP modules (PreMetal1, PreMetal2, PreMetal3, PreTopMetal1), which will increase the thickness variation. If we consider the whole BEOL with the planarized TopMetal1 topography, a typical thickness variation of the $SiO_2$ cladding of around 330 nm results. In this context, the adoption of Metal3 as a mirror below the grating couplers is not only to increase the coupling efficiency, but – more important – to reduce the influence of the $SiO_2$ thickness below the grating. For a $SiO_2$ thickness between a-Si:H and Metal3 of around 3 µm, a thickness variation of around 210 nm can be expected. A deviation in that range is acceptable for the 2D GCs, which will be shown in the next section.

A typical transmitter configuration requires integrated photodetectors for setting the operation point of the modulator.

In the case of a BEOL integrated modulator, an interlayer coupling scheme such as in Ref. [19] will be needed to access a frontend photodiode. This paper will focus on questions related to the polarization multiplexing grating couplers. Therefore, the interlayer coupling is not shown in Fig. 1.

It should be noted that the here reported a-Si:H test structures are fabricated at the Metal2 level instead of the TopMetal1 level, because of some constraints on that test field, given by other test structures. Amongst others, the lower vertical distance to the FOEL is better suitable for initial tests of interlayer coupling structures. Nevertheless, the relevant a-Si:H properties remain the same, independent of whether the integration takes place at Metal2 or at TopMetal1.

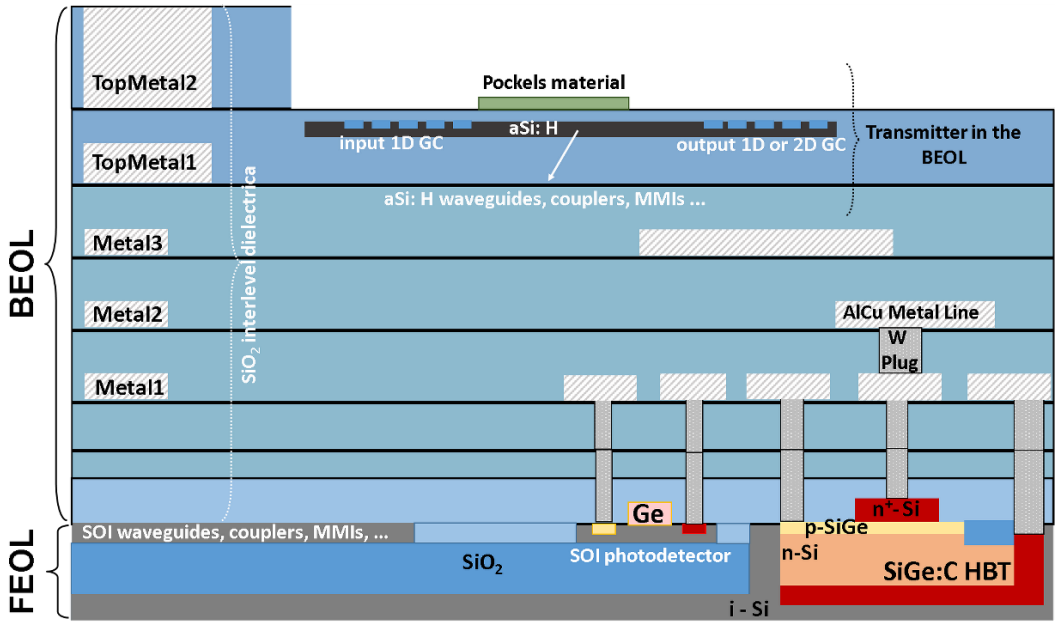

Figure 1: Schematic representation of a 3D photonic BiCMOS platform with electronic and photonic components in the frontend of line (FEOL) and extended by a photonic amorphous Si:H (a-Si:H) layer at the TopMetal1 level. Grating couplers can be used as in- and out-coupling interfaces, which may or may not use Metal3 as a back-reflector. A 2D grating coupler is needed for a polarization-multiplexed system. Pockels material can be bonded for the realization of high-speed modulators and transmitters in the backend of line (BEOL). Realization on both silicon-on-insulator (SOI) and bulk Si wafers is possible.

## 3 Numerical analysis

This section summarizes the numerical results on the first generation a-Si:H 2D GCs at the Metal2 level of our photonic BiCMOS platform and gives further details on the next-generation structures, which will be realized at the TopMetal1 level.

3.1 C-Band a-Si:H 2D GCs at the Metal2 level

The 2D GC's simulation is based on the following assumptions:
- a-Si:H refractive index at 1550 nm: 3.57,
- a-Si:H waveguide height: 220 nm,
- $SiO_2$ thickness between the a-Si:H waveguide and the bulk Si substrate: 3.63 µm,
- $SiO_2$ top cladding: 3 µm.

With these values, the 2D GC design for 1550 nm is developed, which has a 9° coupling angle $\vartheta$ at the symmetry plane between the 2D GC arms ($\varphi = 45°$). To achieve this, we use non-zero angle between the waveguide and the grating, which we call a shear angle. The shear angle is realized by tilting the waveguides with respect to the grating plane. The combination of shear angle and grating period results in a different coupling angle $\vartheta$ at the symmetry plane, as we have explained previously [28]. For the current design, a waveguide-to-grating shear angle of 2° and a grating period of 593 nm are chosen. The diffracting elements are circular with a diameter of 400 nm and an etch depth of 100 nm. The same etch depth is used to define a-Si:H rib waveguides, so that waveguide and 2D GC will be defined in the same etch step.

The numerical analysis is performed by the commercial time-domain finite-integration-technique solver by Simulia CST. The 2D GC's coupling efficiency is calculated as the product of the grating out-coupled power and the mode field overlap with the standard single-mode fiber (SMF) mode with a mode field diameter MFD = 10.4 µm @1550 nm. Fig. 2 shows the simulated coupling efficiency of the designed 2D GC, when any of the two 2D GC waveguides is excited separately. At 1550 nm, a coupling efficiency of -5 dB is expected. At the same wavelength, the cross-polarization, which is the part of a given input polarization converted to its orthogonally polarized counterpart [29], [30] is around -16 dB. The return loss is about 2.3 % (-16.4 dB) at 1550 nm and is caused not only by Fresnel reflection, but also by backscattering of the incident field by the diffracting elements (cf. Ref. [29]). Reduction of the latter effect can reduce the return loss. Finally, with the procedure outlined in Ref. [31], we found that the out-coupled 2D GC signals from the two arms are 95°-polarized to each other at 1550 nm.

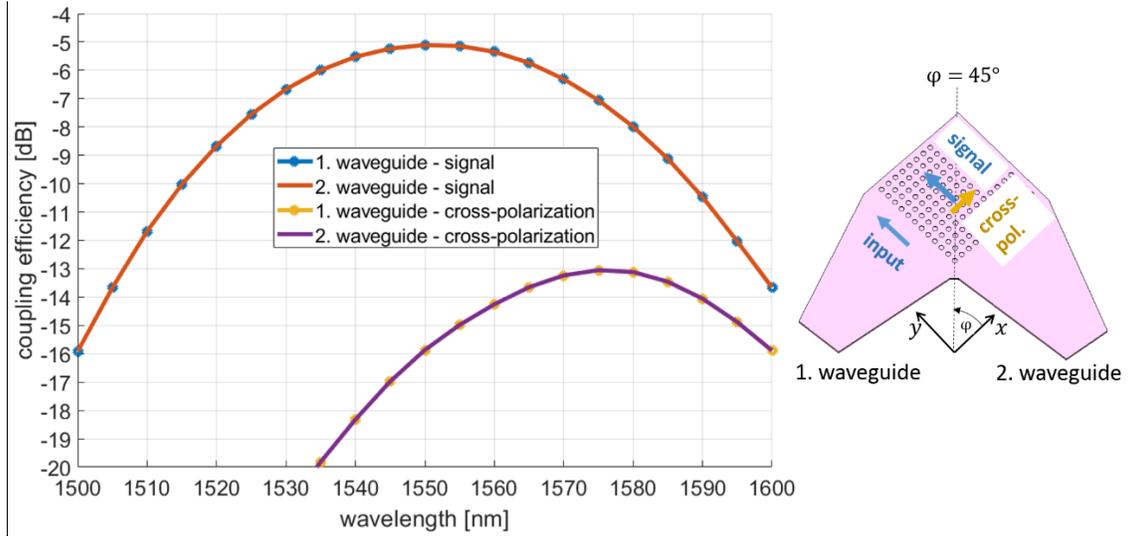

Figure 2: a-Si:H 2D GC design at Metal2 - coupling efficiency of signal and cross-polarization, when either of its two waveguides 1 or 2 is excited. The coupling angle at the symmetry plane is 9°.

Next, we perform an analysis on the coupling efficiency penalties, due to deviations from the optimal coupling position. The considered scenarios account for various tolerances after packaging. The considered alignment deviations are: (1) angular deviation - of the coupling angle $\vartheta$ or of the in-plane angle $\varphi$ and (2) axial displacement along a given direction (on example here - along the *y*-axis). The numerical results show that an angular variation $\Delta\vartheta$ leads to the well-known GC's spectral shift of about 5 nm per 1°. Fig. 3 shows the coupling spectra for the remaining variants: deviation from the symmetry axis $\varphi = 45°$ of (a) $\Delta\varphi = +5°$, (b) $\Delta\varphi = -5°$; displacement along the *y*-axis (c) $\Delta y = -3$ µm and (d) $\Delta y = +3$ µm. The black point gives the SMF central position, while the black arrow assigns the SMF in-plane orientation. Following penalties can be observed:
- For $\varphi = 45° \pm 5°$: the central wavelength of both polarization differs by ~10 nm. The maximum coupling efficiency is not equal. At 1550 nm, the imbalance is small: ~0.2 dB.
- For $\Delta y = -3$ µm: waveguide 2 is favored, leading to 0.8 dB imbalance at 1550 nm.

- For $\Delta y = +3$ µm: the position is non-optimal for both waveguides, the coupling efficiency of both polarizations decreases by 0.6 dB at 1550 nm. Because the deviation is not symmetric, the coupling spectra have different bandwidths.

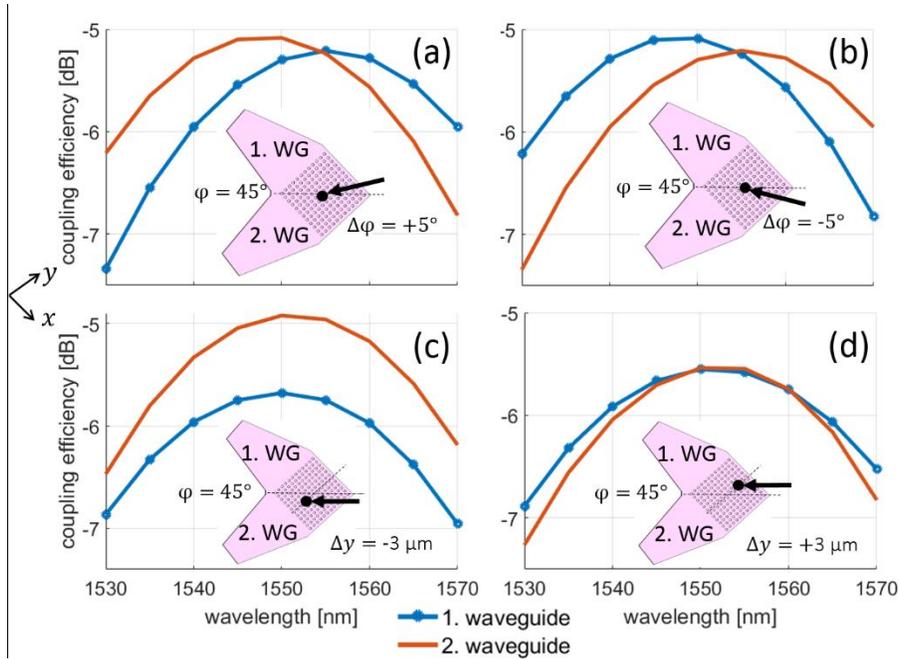

Figure 3: a-Si:H 2D GC design at Metal2 – signal coupling efficiency, when either of its two waveguides (WG) 1 or 2 is excited. The coupling scenarios consider the following SMF deviations from the optimal position: (a) symmetry plane deviation by Δφ = +5°, (b) symmetry plane deviation by Δφ = -5°; (c) displacement along the y-axis Δy = -3 µm and (d) displacement along the y-axis Δy = +3 µm.

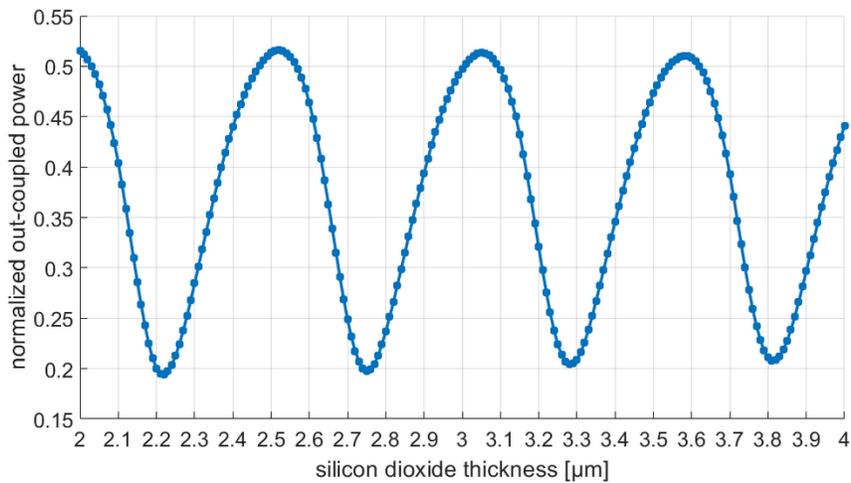

Figure 4: a-Si:H 2D GC design – dependence of the normalized out-coupled power on the $SiO_2$ thickness separation from the Si substrate. The wavelength is 1550 nm.

Next, we analyze another important factor for the performance of a 2D GC, which is the $SiO_2$ thickness between the substrate and the grating. The 2D GC's out-coupled power has a periodic behavior depending on the $SiO_2$ thickness. Out of the periodicity, we are able to estimate an appropriate target value for the future 2D GC realization at the TopMetal1 level. In addition, we need to keep in mind that we are not able to fix its value exacter than in an interval of about ± 100 nm. The chosen target value needs to be robust against variations in this range. Fig. 4 shows the normalized out-coupled power of a 2D GC, when the $SiO_2$ thickness below the grating varies between 2 µm and 4 µm. We can see that the current $SiO_2$ thickness of 3.63 µm is a good value, however not well centered so that within a ± 100 nm interval, the out-coupled power can drop of up to 10 %. This means that our current 2D GC design can have a higher efficiency variation than in the optimal case.

The maximum transmission repeats every 530 nm, with exemplary maximum values of 2 µm, 2.53 µm, 3.06 µm etc. Around these values, a drop of 5 % of the out-coupled power occurs within ± 100 nm, which shows that the mentioned $SiO_2$ thicknesses are good target values for future 2D GC designs. With the boundary conditions of our platform, an appropriate $SiO_2$ thickness at the TopMetal1 level and with Metal3 as a reflector is 3.06 µm. A reference simulation with a bulk Si substrate below the grating is carried out as well. The $SiO_2$ thickness estimated from Fig. 4 is 8.33 µm in that case.

3.2 C-Band and O-Band a-Si:H 2D GCs at theTopMetal1 level

Next, we compare 2D GCs at the TopMetal1 level in the two scenarios: bulk Si below the grating with 8.33 µm $SiO_2$ thickness separation and Metal3 below the grating with 3.06 µm $SiO_2$ thickness separation. In addition, designs for O-band are considered as well. The simulations predict the behavior of future-generation a-Si:H 2D GCs.

There are two reasons why we prefer to use Metal3 as a mirror (instead of Metal1 or Metal2). The first one is its better conductivity resulting in a better reflectivity. The conductivity is calculated from the measured mean value of its sheet resistance, which is 55 mΩ. The second reason is that the choice of Metal3 as a reflector ensures a better $SiO_2$ thickness stability below the grating, which remains in the permitted range of ± 100 nm.

Fig. 5 shows the simulation results for the C-Band a-Si:H 2D GC. Its geometric details remain unchanged. In the bulk Si case, the maximal coupling efficiency remains the same, compared to the structure at the Metal2 level. We obtain -5 dB at 1550 nm. The cross-polarization is slightly higher and reaches -14.4 dB at the same wavelength. This leads to a meager change of the orthogonality relation between the signals from waveguide 1 and 2, which is now 96°. The return loss at 1550 nm is slightly enhanced, becoming 2.6 % (-15.9 dB). If we use the Metal3 as a back-reflector, both signal and cross-polarization are enhanced. The improvement is better pronounced for the signal than for the cross-polarization. With a back-reflector, we obtain a coupling efficiency of -3.4 dB at 1550 nm and a corresponding cross-polarization of -13.4 dB. The return loss is 2.8 % (-15.5 dB). The orthogonality relation is also significantly improved, becoming 89.6°. Among with the efficiency, the signals' orthogonality is dependent on the $SiO_2$ thickness below the 2D GC as well. This can be caused by differences in the $SiO_2$ dependence of signal and cross-polarization.

In the final numerical analysis, we compare similarly 2D GCs with bulk Si or metal mirror in the O-Band. Coherent solutions using O-band have the potential to compete in the data center domain as well [32]. We choose the same $SiO_2$ thicknesses as for C-band, which are not necessarily optimal for O-band. Fig. 6 shows the simulation results for a-Si:H O-band 2D GCs with a waveguide-to-grating shear angle of 2°, a grating period of 458 nm, circular diffracting elements with a diameter of 280 nm and an etch depth of 100 nm. The calculation assumes a SMF MFD of 9.2 µm. In both the bulk Si and the metal mirror cases, we observe a better coupling efficiency and a lower cross-polarization than those of the C-band 2D GC. With bulk Si below, the O-Band 2D GC has a maximal coupling efficiency of -4 dB at 1300 nm and a cross-polarization of -18.6 dB. The return loss is about 1.9% (-17.2 dB). With the back-reflector Metal3, an improvement of wavelengths shorter than 1300 nm is better pronounced, so that the maximum looks shifted towards 1295 nm. At 1300 nm, we obtain a coupling efficiency of -2.8 dB, at 1295 nm it is -2.7 dB. The back-reflector enhances the cross-polarization, except in a small wavelength range from 1300 nm to 1310 nm. At 1300 nm we have a cross-polarization of -18.8 dB. The return loss increases to 3 % (-15.2 dB). Regarding the orthogonality relation of the signals from the two GC arms, at 1300 nm we obtain 92.7° in the bulk Si case and 91.8° with Metal3 below. In our platform, the integration of O-band structures at the TopMetal1 level can be more advantageous in comparison to C-band, provided that the waveguide loss is of similar order.

Discussing these results, we need to point out that the metals available in our standard BiCMOS process do not have very high reflectivity such as e.g. to gold. The achievable coupling efficiency is thus lower than that of reported 2D GCs with bonded gold mirrors [33], [34]. Furthermore, the metal thickness is a parameter that can vary. The conductivity may be influenced by material impurities as well. Deviations due to the latter effect are part of the specified metal sheet resistance's variation, which is below 3 % per specification. Another issue is the metal roughness, which is, however, lower for thin metals. These aspects may be responsible for an inferior improvement of the coupling efficiency than expected. Their impact on the efficiency's variation needs to be investigated in more detail in future.

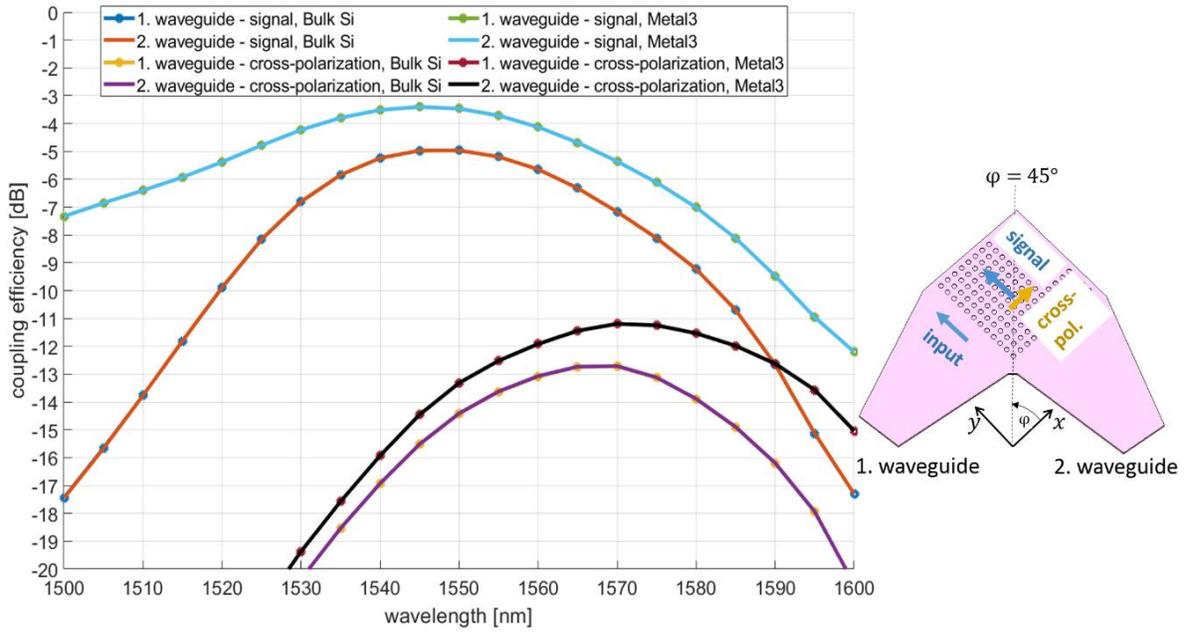

Figure 5: Coupling efficiency of signal and cross-polarization for an a-Si:H 2D GC for C-Band at the TopMetal1 level. Two different back-reflectors are considered: bulk Si at a distance of 8.33 μm and Metal3 at a distance of 3.06 μm.

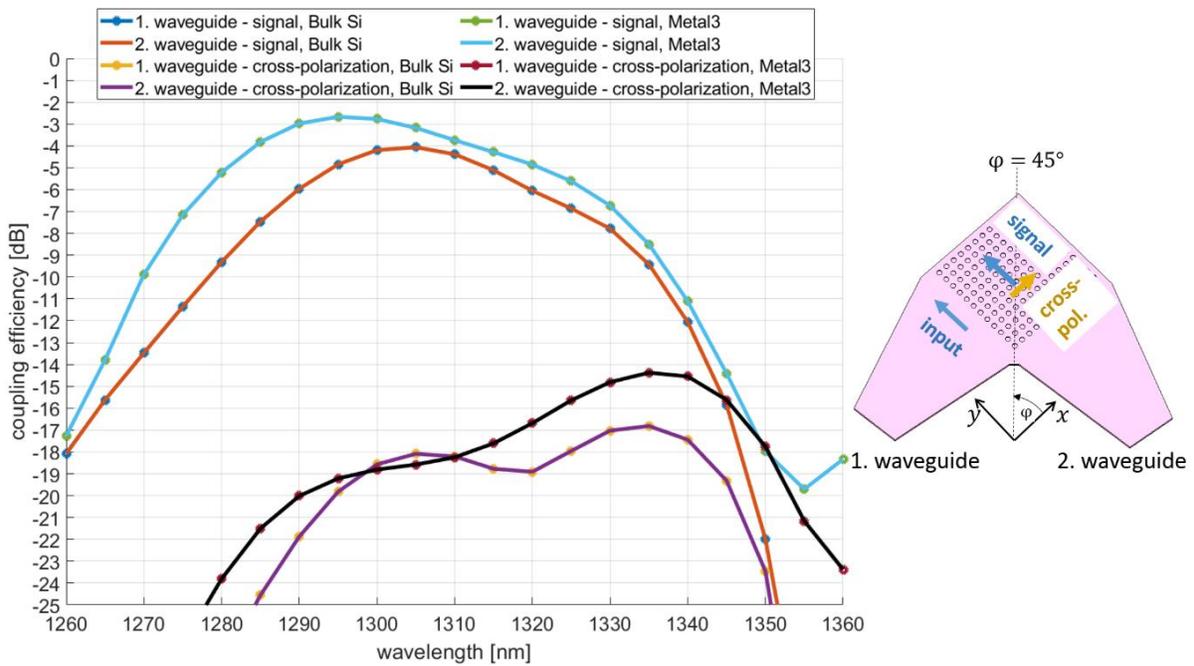

Figure 6 Coupling efficiency of signal and cross-polarization for an a-Si:H 2D GC for O-Band at the TopMetal1 level. Two different back-reflectors are considered: bulk Si at a distance of 8.33 μm and Metal3 at a distance of 3.06 μm.

## 4 Fabrication and process evaluation

In this section, general information about the BEOL a-Si:H fabrication flow is given. Furthermore, process evaluation with a focus on the waveguide uniformity is carried out.

4.1 Fabrication process of a-Si:H waveguides

For the fabrication and the integration of a-Si:H waveguides in the BEOL of the photonic BiCMOS technology, there are certain limitations due to the risk of metal contamination of tools used for the FEOL. Generally, independent in which level of the BEOL the a-Si:H waveguides shall be integrated the first step is a CMP procedure of $SiO_2$ to planarize the metal topography. The tool for $SiO_2$ CMP is also used for the shallow trench isolation in the FEOL. Due to the roughness of the metal surface, it is extremely risky to polish $SiO_2$ very close to the metal layer in the BEOL. There is a high probability to polish also the metal surface, which would result in a metal contamination of the CMP tool.

For the realization of the a-Si waveguides in the BEOL, it is not allowed to use any cleaning steps such as those we use for the production of our SOI waveguides in the FEOL. This is because the necessary tools are also used for pre cleaning steps before any epitaxial process. This means that it would be very probable to contaminate the epitaxy tool with metal particles. Therefore, for the first tests we start two parallel process flows to develop a BEOL compatible integration process for the a-Si:H waveguides.

In the first iteration, to fabricate a-Si:H waveguides, we use 200 mm silicon test wafers as a substrate. We start with the deposition of 3630 nm $SiO_2$ and 220 nm a-Si by plasma enhanced chemical vapor deposition (PECVD). For the a-Si:H waveguide test structures, we have to process three layers. Fig. 7 shows a scheme of the process flow. At first, we structure the 1D GCs by reactive ion etching (RIE) over a resist mask. The realization of waveguide ribs/ 2D GCs and deep etched waveguides is carried out in the same way. Because we need different etch depths for 1D GCs (70 nm), waveguide ribs/ 2D GCs (100 nm) and deep etched waveguides (220 nm), they need to be processed separately. At the end, we deposit a thick $SiO_2$ by PECVD and planarize the waveguide topography by CMP. As a result, we obtain a 3 µm $SiO_2$ cladding above the grating couplers.

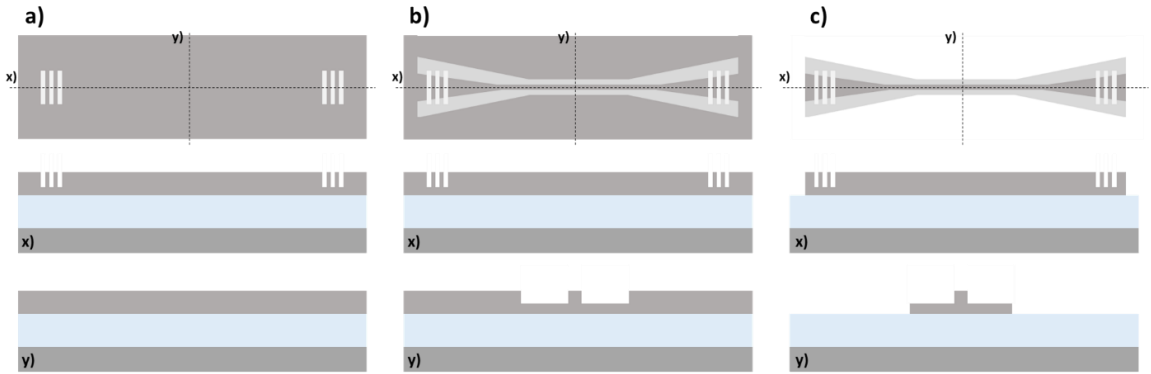

Figure 7: Process flow - after deposition of 3630 nm $SiO_2$ and 220 nm a-Si, 1D grating couplers (a), waveguide ribs (b), and deep etched waveguides (c) are fabricated.

4.2 Uniformity of a-Si:H waveguides

The uniformity of BEOL integrated a-Si:H waveguides is an important aspect, since many integrated devices including 1D and 2D GCs have a strong dependence on the mode effective refractive index. Here, two main sources of its variation are analyzed – waveguide height deviation and variance of the a-Si:H refractive index. In Table 1 the thickness at different wafer positions of the a-Si:H after its deposition are summarized.

The variation of the 220 nm a-Si:H waveguide thickness is < 4 % over the wafer. This is still considerably larger than the thickness variation of 220 nm crystalline SOI, which is < 1%. At present, further optimization of the deposition process is undergoing to understand the reachable limits of the a-Si:H layer thickness variation.

Next, wafer-level uniformity of refractive index of the deposited material is investigated. For the determination of the a-Si:H refractive index at 1550 nm, ellipsometry measurements on 2 wafers are carried out. On the first wafer, the mean refractive index is 3.6082 and the minimum-maximum range is 0.0165. On the second wafer, the mean refractive index is again 3.6082 with a lower range of 0.0145. These results correspond well with typical 200 mm tool specifications.

Table 1: Measurement results of amorphous silicon thickness wafer distribution and range.

| Chip No. | a-Si:H layer thickness [nm] |
|---|---|
| 1 | 226.3 |
| 2 | 226.1 |
| 3 | 220.8 |
| 4 | 219.3 |
| 5 | 221.2 |
| 6 | 223.9 |
| 7 | 223 |
| 8 | 222.5 |
| 9 | 217.6 |
| Mean | 222.3 |
| Min | 217.6 |
| Max | 226.3 |
| Range | 8.6 |

## 5 Experimental evaluation of a-Si:H waveguides and 2D GCs in the BEOL

In this section, first experimental results on a-Si:H waveguide loss and coupling efficiency are summarized. All structures are realized at the Metal2 level of the photonic BiCMOS platform used.

5.1 Linear waveguide loss

The total waveguide loss analyzed here can be attributed to sidewall scattering and material absorption. We use a rib waveguide with a core width of 500 nm, a core height of 220 nm and a rib etch depth of 100 nm. The waveguide loss, is measured automatically on wafer-scale with the optical backscatter reflectometry method [35]. The loss is averaged over 59 dies and in a wavelength range from 1526 nm to 1566 nm. Fig. 8 shows a wafer-map and a histogram of the waveguide loss distribution with a mean value of 6.2 dB/cm and a 3σ interval of ±1.5 dB/cm, where σ is the standard deviation. In comparison, deeply etched (120 nm) SOI rib waveguides on our platform have a typical loss of about 3 dB/cm per specification.

5.2 Coupling efficiency

Due to the larger refractive index of the fabricated a-Si:H, the 2D GC maximum efficiency is shifted towards the L-band. Measurement results are evaluated at a 1580 nm wavelength. The setup consists of a tunable laser Agilent 81940A, a manual polarization controller and two SMFs for the in- and out-coupling. The signal is detected with a power meter Agilent 81634B. The chip input power is determined by a calibrated photodiode Thorlabs SM05PD5A. A manual wafer prober is used for wafer-scale measurements. Fig. 9 shows a schematic of the test structures used for the characterization of the gratings coupling efficiency. The structures shown are only illustrative and do not represent devices from this experiment. For test purposes, we designed a 1D GC with a period of 610 nm, a slot width of 315 nm and an etch depth of 70 nm. The measurement of the coupling efficiency of 1D and 2D GCs is performed in the following way. First, back-to-back measurements at 13° of the structure in Fig. 9 (a) are performed, out of which we are able to calculate the 1D GC's coupling spectrum. Next, we measure the coupling spectrum of the structure in Fig. 9 (b). We use the 1D GC side as an input and the 2D GC as an output, in order to consider the 2D GC in a transmitter-side configuration. In this case, we have two clearly defined on-chip polarization states at the 2D GC. For symmetry reasons, we show the results only when light was coupled at the input In 1.

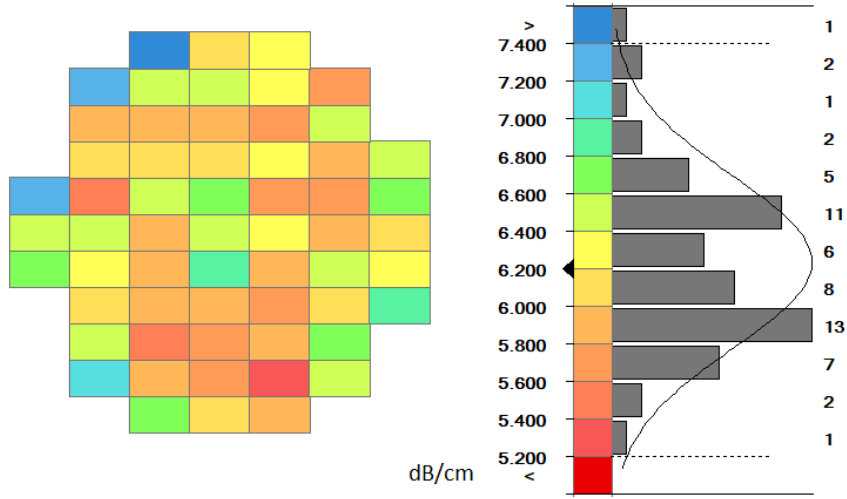

Figure 8: Wafer map and histogram of the loss distribution of an a-Si:H rib waveguide with a core width of 500 nm, core height of 220 nm and rib etch depth of 100 nm. The mean loss ± 3σ is 6.2 dB/cm ± 1.5 dB/cm.

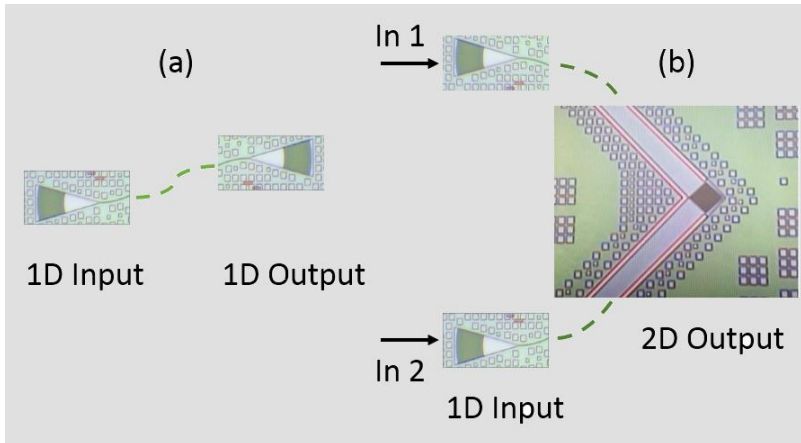

Figure 9: Schematic of the test structures for the determination of the a-Si:H 1D and 2D GC efficiency. (a) 1D-1D configuration, (b) 1D-2D configuration, in which the 1D GC acts an input interface. The depicted structures are only illustrative and do not show the actual devices.

To obtain the 2D GC spectrum, the 1D GC spectrum from the 1D-1D GC measurement is subtracted. For an appropriate 2D GC characterization, the 1D and 2D GC spectra must be centered at the same wavelength. For that reason, the coupling angles in the 1D-2D configuration are 13° at the 1D input and 8° at the 2D output. The coupling efficiency is averaged over 17 out of 59 chips on the wafer. Several exemplary spectra are shown in Fig. 10 (a) for 1D GCs and (b) for 2D GCs. At 1580 nm, we find a 1D GC efficiency of -3.6 dB with a 3σ interval of ±0.6 dB. The 2D GCs have a mean coupling efficiency ±3σ of -5 dB ±1.2 dB. With this, 2D GC efficiency is in excellent agreement with the simulation results. However, the coupling efficiency has a significant variation on the wafer, which is evident from the comparison of Fig. 10 (a) and (b). The variation can be attributed to the non-optimal target $SiO_2$ thickness below the 2D GCs, which is not sufficiently robust against variations within ± 100 nm. Future designs will target at a more appropriate SiO$_2$ thickness to avoid the strong influence of its deviation.

Finally, we compare the a-Si:H 1D and 2D GCs with the same couplers realized on SOI with a crystalline Si (c-Si) waveguide height of 220 nm and a buried oxide (BOX) thickness of 2 µm. The cladding thickness is the same as for the a-Si:H structures. Here, 9 dies on the wafer are considered. The statistics for the a-Si:H GCs at the same positions do not change significantly from the values reported in the previous paragraph. Fig. 11 (a) and (b) show exemplary coupling spectra for the 1D and 2D GCs in c-Si. Because the refractive index of c-Si is much lower (3.47 vs. 3.63 of a-Si:H), the spectra for the same coupling angles are shifted towards a 1530 nm wavelength. For the 1D GCs, the mean coupling efficiency ±3σ is -3.8 dB ±1 dB at 1530 nm. For the 2D GCs, we obtain -4.4 dB ± 0.6 dB. With this, the coupling efficiencies in both 1D and 2D case do not differ significantly from their a-Si:H counterparts. Differences of less than 0.5 dB are difficult to dissolve due to the limited mechanical precision of our equipment. In the SOI case, we observe this time less 2D GC efficiency variation. The reason can be that the BOX thickness is more appropriate with respect to robustness of 2D GCs against variations.

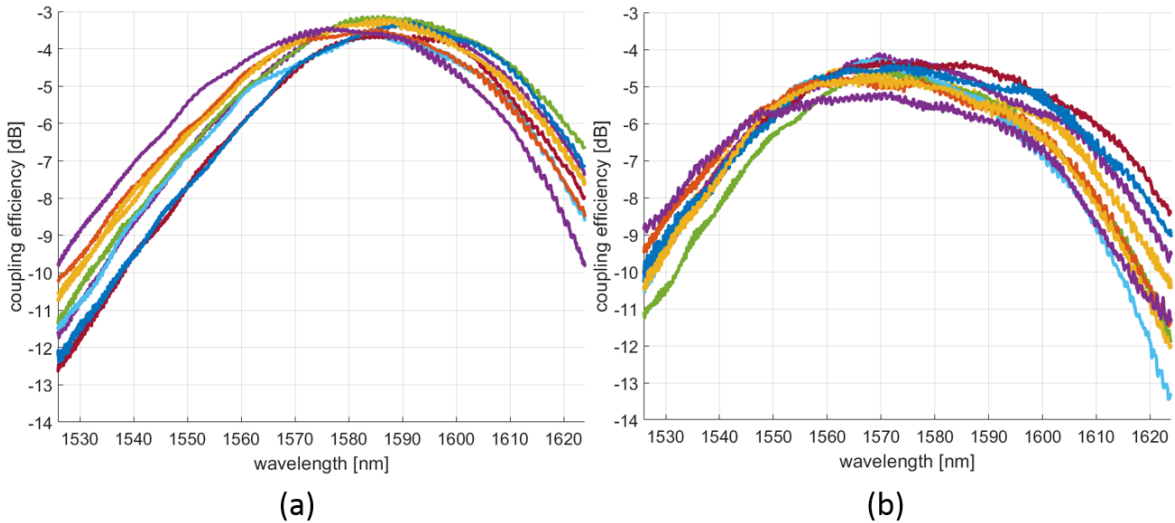

Figure 10: Exemplary coupling spectra, measured at different wafer dies. (a) a-Si:H 1D GCs, (b) a-Si:H 2D GCs.

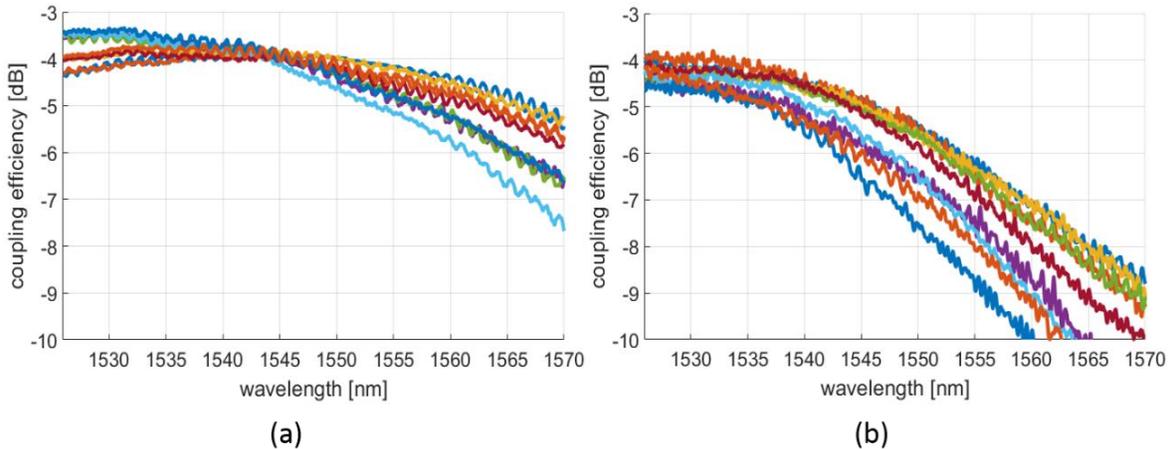

Figure 11: Exemplary coupling spectra, measured at different SOI wafer dies. (a) c-Si 1D GCs, (b) c-Si 2D GCs.

## 4 Conclusions

In this work, we analyze the first generation of a-Si:H waveguides and grating couplers, fabricated in a photonic BiCMOS BEOL. In terms of linear loss, our standard c-Si waveguides out-perform the first fabricated a-Si:H waveguides. In Ref. [16] low-loss a-Si:H rib waveguides were realized by CMP-planarization of the waveguide top surface. In future work, this approach could be pursued for loss reduction of our waveguides.

We analyze in more detail first a-Si:H 2D grating couplers, realized at the Metal2 level. Experimentally, a mean loss ±3σ of -5 dB ± 1.2 dB is obtained. Compared to c-Si 2D grating couplers with a mean coupling loss ±3σ of -4.4 dB ± 0.6 dB, a-Si:H 2D grating couplers show a similar performance in terms of average coupling efficiency. The higher variation is caused on the one hand by larger variation of the a-Si:H waveguide thickness. Presently, the reachable minimum of a-Si:H waveguide thickness variation is under investigation. Another aspect is the target $SiO_2$ thickness below the a-Si:H 2D GCs, which is in this case not optimal with respect to robustness against $SiO_2$ thickness deviations. We determined appropriate target $SiO_2$ thicknesses for future designs, for which a variation within ± 100 nm will not interfere the 2D grating coupler's performance. In future, a-Si:H 2D grating couplers need to be integrated at the TopMetal1 level of our photonic BiCMOS platform. The usage of Metal3 3.06 µm below the grating could bring more advantages for two reasons. The first one is the $SiO_2$ thickness variation, which will be reduced, compared to the bulk Si case. The second one are the good properties of Metal3 in terms of conductivity and surface roughness, which make it suitable as a grating back-reflector. Simulations predict a coupling efficiency of -3.4 dB for C-Band, which is 1.6 dB better than in the case of bulk Si below the gratings. If we consider extension towards O-band, the 2D GCs can reach -2.7 dB with Metal3 as a back-reflector. The here reported performance evaluation of a-Si:H waveguides and 2D GCs indicate the feasibility of a-Si:H based BEOL integration approach of coherent transmitters, notwithstanding the requirement of further process optimization.

**Acknowledgements**

This work was supported in part by the German Research Foundation (DFG) through the projects EPIC-Sense (ZI 1283-6-1) and EPIDAC (ZI 1283-7-1), by the Federal Ministry of Education and Research (BMBF) through project PEARLS (13N14932).

**References**

[1] X. Zhou, R. Urata and H. Liu, "Beyond 1 Tb/s Intra-Data Center Interconnect Technology: IM-DD or Coherent?," *Journal of Lightwave Technology,* vol. 38, no. 2, p. 475–484, 2020.

[2] NeoPhotonics, *White Paper - Photonic IC Enabled Coherent Optical Systems.*

[3] J. Zhou, J. Wang, L. Zhu and Q. Zhang, "Silicon Photonics for 100 Gbaud," *Journal of Lightwave Technology,* vol. 39, no. 4, p. 857–867, 2021.

[4] A. Mekis, S. Gloeckner, G. Masini, A. Narasimha, T. Pinguet, S. Sahni and P. D. Dobbelaere, "A Grating-Coupler-Enabled CMOS Photonics Platform," *IEEE Journal of Selected Topics in Quantum Electronics,* vol. 17, no. 3, p. 597–608, 2011.

[5] L. Zimmermann, D. Knoll, M. Kroh, S. Lischke, D. Petousi, G. Winzer and Y. Yamamoto, "BiCMOS Silicon Photonics Platform," in *Optical Fiber Communication Conference (OFC) 2015*, Los Angeles, CA, USA, 2015.

[6] M. Rakowski, C. Meagher, K. Nummy, A. Aboketaf, J. Ayala, Y. Bian, B. Harris, K. Mclean, K. McStay, A. Sahin, L. Medina, B. Peng, Z. Sowinski, A. Stricker, T. Houghton, C. Hedges, K. Giewont, A. Jacob, T. Letavic, D. Riggs, A. Yu and J. Pellerin, "45nm CMOS - Silicon Photonics Monolithic Technology (45CLO) for next-generation, low power and high speed optical interconnects," in *Optical Fiber Communication Conference (OFC) 2020*, San Diego, CA, USA, 2020.

[7] X. Zhou, R. Urata and H. Liu, "Beyond 1Tb/s Datacenter Interconnect Technology: Challenges and Solutions (Invited)," in *Optical Fiber Communication Conference (OFC) 2019*, San Diego, CA, USA, 2019.

[8] D. Petousi, L. Zimmermann, K. Voigt and K. Petermann, "Performance Limits of Depletion-Type Silicon Mach-Zehnder Modulators for Telecom Applications," *Journal of Lightwave Technology,* vol. 31, no. 22, p. 3556–3562, 2013.

[9] D. Petousi, L. Zimmermann, A. Gajda, M. Kroh, K. Voigt, G. Winzer, B. Tillack and K. Petermann, "Analysis of Optical and Electrical Tradeoffs of Traveling-Wave Depletion-Type Si Mach-Zehnder Modulators for High-

Speed Operation," *IEEE Journal of Selected Topics in Quantum Electronics,* vol. 21, no. 4, p. 199–206, 2015.

[10] K. Alexander, J. P. George, J. Verbist, K. Neyts, B. Kuyken, D. V. Thourhout and J. Beeckman, "Nanophotonic Pockels modulators on a silicon nitride platform," *Nature Communications,* vol. 9, no. 1, pp. 1-6, 2018.

[11] F. Eltes, C. Mai, D. Caimi, M. Kroh, Y. Popoff, G. Winzer, D. Petousi, S. Lischke, J. E. Ortmann, L. Czornomaz, L. Zimmermann, J. Fompeyrine and S. Abel, "A BaTiO3-Based Electro-Optic Pockels Modulator Monolithically Integrated on an Advanced Silicon Photonics Platform," *Journal of Lightwave Technology,* vol. 37, no. 5, p. 1456–1462, 2019.

[12] A. Rao, A. Patil, P. Rabiei, A. Honardoost, R. DeSalvo, A. Paolella and S. Fathpour, "High-performance and linear thin-film lithium niobate Mach-Zehnder modulators on silicon up to 50 GHz," *Optics Letters,* vol. 41, no. 24, pp. 5700-5703, 2016.

[13] P. O. Weigel, J. Zhao, K. Fang, H. Al-Rubaye, D. Trotter, D. Hood, J. Mudrick, C. Dallo, A. T. Pomerene, A. L. Starbuck, C. T. DeRose, A. L. Lentine, G. Rebeiz and S. Mookherjea, "Bonded thin film lithium niobate modulator on a silicon photonics platform exceeding 100 GHz 3-dB electrical modulation bandwidth," *Optics Express,* vol. 26, no. 18, pp. 23728-23739, 2018.

[14] N. Boynton, H. Cai, M. Gehl, S. Arterburn, C. Dallo, A. Pomerene, A. Starbuck, D. Hood, D. C. Trotter, T. Friedmann, C. T. DeRose and A. Lentine, "A heterogeneously integrated silicon photonic/lithium niobate travelling wave electro-optic modulator," *Optics Express,* vol. 28, no. 2, pp. 1868-1884, 2020.

[15] R. Takei, Y. Maegami, E. Omoda, Y. Sakakibara, M. Mori and T. Kamei, "Low-loss and low wavelength-dependence vertical interlayer transition for 3D silicon photonics," *Optics Express,* vol. 23, no. 14, pp. 18602-18610, 2015.

[16] R. Takei, S. Manako, E. Omoda, Y. Sakakibara, M. Mori and T. Kamei, "Sub-1 dB/cm submicrometer-scale amorphous silicon waveguide for backend on-chip optical interconnect," *Optics Express,* vol. 22, no. 4, pp. 4779-4788, 2014.

[17] N. Sherwood-Droz and M. Lipson, "Scalable 3D dense integration of photonics on bulk silicon," *Optics Express,* vol. 19, no. 18, pp. 17758-17765, 2011.

[18] E. W. Ong, N. M. Fahrenkopf and D. D. Coolbaugh, "SiNx bilayer grating coupler for photonic systems," *OSA Continuum,* vol. 1, no. 1, pp. 13-25, 2018.

[19] S. Zhu and G.-Q. Lo, "Vertically Stacked Multilayer Photonics on Bulk Silicon Toward Three-Dimensional Integration," *Journal of Lightwave Technology,* vol. 34, no. 2, p. 386–392, 2016.

[20] J. H. Kang, Y. Atsumi, Y. Hayashi, J. Suzuki, Y. Kuno, T. Amemiya, N. Nishiyama and S. Arai, "Amorphous-Silicon Inter-Layer Grating Couplers With Metal Mirrors Toward 3-D Interconnection," *IEEE Journal of Selected Topics in Quantum Electronics,* vol. 20, no. 4, p. 317–322, 2014.

[21] R. Takei, "Amorphous Silicon Photonics," in *Crystalline and Non-crystalline Solids*, InTech, 2016.

[22] D. Knoll, S. Lischke, R. Barth, L. Zimmermann, B. Heinemann, H. Rücker, C. Mai, M. Kroh, A. Peczek, A. Awny, C. Ulusoy, A. Trusch, A. Kruger, J. Drews, M. Fraschke, D. Schmidt, M. Lisker, K. Voigt, E. Krune and A. Mai, "High-performance photonic BiCMOS process for the fabrication of high-bandwidth electronic-photonic integrated circuits," in *2015 IEEE International Electron Devices Meeting (IEDM)*, Washington, DC, USA, 2015.

[23] D. Knoll, S. Lischke, A. Awny, M. Kroh, E. Krune, C. Mai, A. Peczek, D. Petousi, S. Simon, K. Voigt, G. Winzer and L. Zimmermann, "BiCMOS silicon photonics platform for fabrication of high-bandwidth electronic-photonic integrated circuits," in *2016 IEEE 16th Topical Meeting on Silicon Monolithic Integrated Circuits in RF Systems (SiRF)*, Austin, TX, USA, 2016.

[24] R. Marchetti, C. Lacava, L. Carroll, K. Gradkowski and P. Minzioni, "Coupling strategies for silicon photonics integrated chips [Invited]," *Photonics Research,* vol. 7, no. 2, pp. 201-239, 2019.

[25] T. Mitze, M. Schnarrenberger, L. Zimmermann, J. Bruns, F. Fidorra, K. Janiak, J. Kreissl, S. Fidorra, H. Heidrich and K. Petermann, "CWDM Transmitter Module Based on Hybrid Integration," *IEEE Journal of selected topics in quantum electronics,* vol. 12, no. 5, pp. 983-987, 2006.

[26] H. Duprez, A. Descos, T. Ferrotti, C. Sciancalepore, C. Jany, K. Hassan, C. Seassal, S. Menezo und B. Bakir, „1310 nm hybrid InP/InGaAsP on silicon distributed feedback laser with high side-mode suppression ratio," *Optics Express,* Bd. 23, Nr. 7, pp. 8489-8497, 2015.

[27] D. Shin, J. Cha, S. Kim, Y. Shin, K. Cho, K. Ha, G. Jeong, H. Hong, K. Lee and H.-K. Kang, "O-band DFB laser heterogeneously integrated on a bulk-silicon platform," *Optics Express,* vol. 26, no. 11, pp. 14768-14774, 2018.

[28] G. Georgieva and K. Petermann, "Analytical and Numerical Investigation of Silicon Photonic 2D Grating Couplers with a Waveguide-to-Grating Shear Angle," in *2018 Progress in Electromagnetics Research Symposium (PIERS)*, Toyama, Japan, 2018.

[29] G. Georgieva, K. Voigt, C. Mai, P. M. Seiler, K. Petermann and L. Zimmermann, "Cross-polarization effects in sheared 2D grating couplers in a photonic BiCMOS technology," *Japanese Journal of Applied Physics,* vol. 59, no. SO, p. SOOB03, 2020.

[30] G. Georgieva, K. Voigt, P. M. Seiler, C. Mai, K. Petermann and L. Zimmermann, "A physical origin of cross-polarization and higher-order modes in two-dimensional (2D) grating couplers and the related device performance limitations," *Journal of Physics: Photonics,* vol. 3, no. 3, p. 035002, 2021.

[31] G. Georgieva, P. M. Seiler, C. Mai, K. Petermann and L. Zimmermann, "2D Grating Coupler Induced Polarization Crosstalk in Coherent Transceivers for Next Generation Data Center Interconnects," in *The Optical Fiber Communication Conference (OFC) 2021*, 2021.

[32] P. M. Seiler, G. Georgieva, G. Winzer, A. Peczek, K. Voigt, S. Lischke, A. Fatemi and L. Zimmermann, "Toward Coherent O-Band Data Center Interconnects," *Frontiers of Optoelectronics,* pp. 1-12, 2021.

[33] L. Carroll, D. Gerace, I. Cristiani, S. Menezo and L. C. Andreani, "Broad parameter optimization of polarization-diversity 2D grating couplers for silicon photonics," *Optics Express,* vol. 21, no. 18, pp. 21556-21568, 2013.

[34] Y. Luo, Z. Nong, S. Gao, H. huang, Y. Zhu, L. Liu, L. Zhou, J. Xu, L. Liu, S. Yu and X. Cai, "Low-loss two-dimensional silicon photonic grating coupler with a backside metal mirror," *Optics Letters,* vol. 43, no. 3, pp. 474-477, 2018.

[35] A. Peczek, C. Mai, G. Winzer and L. Zimmermann, "Comparison of cut-back method and optical backscatter reflectometry for wafer level waveguide characterization," in *2020 IEEE 33rd International Conference on Microelectronic Test Structures (ICMTS)*, Edinburgh, UK, 2020.